# Ground Moving Target Detection Using Multi-Features under Antenna Array Crabbing

Rafi Ahmed[(1)] and Hai Deng[(1)]

(1) Department of Electrical and Computer Engineering, Florida International University, Miami, FL 33174, USA

rahme007@fiu.edu, hai.deng@fiu.edu

*Abstract*—This paper introduces a new moving target detection method under crabbing error due to platform motion. Two contour-based attributes known as circularity ratio and bending energy are extracted to a transformed domain of the radar samples. These features are applied to detect the target and non-linear clutter using a machine learning classifier. The effectiveness of the contour-based features is validated with simulation results.

## I. INTRODUCTION

In radar applications, ground clutter cancellation becomes very difficult due to the misalignment of antenna array and platform velocity [1]. Especially in antenna arrays, this problem causes severe performance deterioration together with the mutual coupling [1,2]. Space-time adaptive processing (STAP) is a very popular solution to suppress this type of clutter. However, the method requires a substantial amount of secondary data for the clutter removal [1,3]. To counter this problem, a proximity feature-based approach is proposed in [4]. However, it suffers difficulty when the target is very close to the clutter. Hence, a more robust detection is needed to detect the target and clutter in this scenario.

This paper proposes a new detection method using two contour-based features such as, circularity ratio and bending energy extracted from the transformed domain of space-time samples. A classifier is trained with these features' data. The trained classifier can then detect the target from the non-linear clutter.

## II. SIGNAL MODEL FOR MOVING TARGET DETECTION

An airborne radar is considered with an array of *N* uniform linear antenna elements. The radar also transmits *M* pulses in one dwell interval. Also, the radar's side antenna array axis is not aligned with the platform velocity. Therefore, an crabbing error angle $\chi$ will be introduced that causes frontlobe to combine with backlobe clutter. As a result, the clutter ridge takes on an elliptical shape instead of a line. Now, the space-time vectors for clutter, thermal noise and target can be expressed as $\mathbf{v_c}$, $\mathbf{v_n}$ and $\mathbf{v_t}$ under alternate ($H_1$) hypothesis respectively. Whereas only clutter and noise samples are present in null ($H_0$) hypothesis. Hence, the covariance matrix for all the components under $H_1$ hypothesis can be written as,

$$\mathbf{K} = \mathbf{K}_c + \mathbf{K}_n + \mathbf{K}_t \qquad (1)$$

Now, the covariance matrix can be converted to Doppler frequency domain via minimum variance (MV) transform. A denoising process is then applied to reduce thermal noise. After the reduction of thermal noise, most of the signal pixels in the angle-Doppler scene are the components of target or clutter only.

## III. CONTOUR BASED TARGET DETECTION METHOD

The denoised scene is now processed for pixel clusters using a flood fill algorithm. The algorithm yields non-zero connected signal pixels of the denoised angle-Doppler domain. Afterwards, circularity ratio and bending energy features are extracted.

### A. Circularity Ratio (CR)

Circularity ratio (CR) is defined as the ratio of the connected components' area to the perimeter of the connected contour. It can be written as,

$$CR = 4\pi A_c P_c^{-1} \qquad (2)$$

Where $A_c$ is the area of the connected region. This area is the number of all non-zero pixels inside that connected region. Whereas the perimeter can be found using the boundary points. The boundary points (n) of a certain connected area can be determined by a tracing algorithm. Hence the perimeter is given by:

$$P_c = \sum_{t=1}^{n-1}\left(\sqrt{(x_{t+1}-x_t)^2 + (y_{t+1}-y_t)^2}\right) + D \qquad (3)$$

Here, *D* is the Euclidean distance between the first and last point of the connected boundary. The value of this ratio increases for target and decreases for clutters. This contour feature is equally applicable for linear clutter detection.

### B. Bending Energy($E_B$)

A connected block in the angle-Doppler space appears to have curvature in its contour edges. Different types of change relating to contour edges can be observed for target and interference. Specifically, when the clutter produces elliptical shape, features like bending energy could provide distinguishing information. Bending energy ($E_B$) determines the change of contour using the boundary edge points from the boundary tracing algorithm. This contour change can be given by:

$$\delta(n) = \frac{\alpha_{n+1} - \alpha_n}{\|\ell_{n+1} - \ell_n\|}; \; n = 1, 2 \ldots N \qquad (4)$$

Where, $\alpha$ is the angle and $\ell$ is the coordinate of the edge points for the curve segment. Therefore, the bending energy can be expressed as,

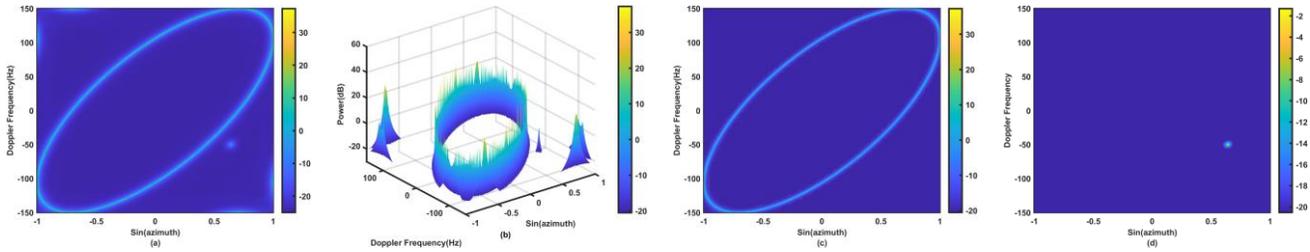

Figure 1. Target and elliptic clutter detection: (a) target at −50 Hz and non-linear clutter with antenna misalignment of 45°.(b) denoised 3D-plot of radar scene; (c) detected clutter and (d) target with contour-based features.

$$E_B = mean(|\delta(n)|^2) \quad (5)$$

Usually, the target has more bending energy than the clutter. This feature is also applicable when target overlaps with clutter.

Using these two features, one can build a classifier to detect the target and non-linear clutter. If **f** is the feature vector of these two features, then the following metric can be defined using Mahalanobis distance:

$$d_s = \sqrt{(\mathbf{f}-\mathbf{m})^T \mathbf{C}^{-1} (\mathbf{f}-\mathbf{m})}; \quad s=0,1 \quad (6)$$

Where **m** is the mean feature vector attained from the training data with $H_s$ hypothesis and **C** is the covariance matrix of the feature vector. **C** is estimated from the training data. If $d_1 > d_0$, the connected region is classified as clutter. Otherwise, it is a target.

## IV. SIMULATION RESULTS

We simulated an airborne radar that transmits 18 coherent pulses during one dwell interval. The radar contains 18 antenna elements with half-wavelength spacing. The operating frequency is 450 MHz and pulse repetition frequency is 300 Hz. The clutter-to-noise-ratio is set at 40 dB. The signal-to-noise-ratio (SNR) is 0 dB. The antenna misalignment error (crab) angle is assumed to be 45°. An MV transformed angle-Doppler radar scene is then simulated with these parameters. The clutter is Gaussian distributed with randomly changing its amplitude. Furthermore, a training dataset containing 2000 samples is generated using Monte Carlo experiment for null and alternate hypotheses. The classifier is then trained using (6) to detect the target and clutter block.

With platform velocity of 50 m/s, a point target is set at −50 Hz of Doppler frequency in the ground. The spatio-temporal data is then transformed to angle-Doppler domain using MV method. Fig. 1(a) shows the angle-Doppler space scenario where clutter forms an elliptical shape. The scene is then denoised with a threshold based on the standard deviation of the bottom part of the angle-Doppler data. Fig 1(b) shows the denoised 3D version of the target and clutter. Here it can be seen that there are some components evident at edges of the left and right corner. These non-zero pixels appear due to the grating lobes of the array spacing. However, they can be disregarded as nontargets since they are at the edges. Afterwards, flood fill algorithm is applied to get the individual connected regions from the angle-Doppler image. The algorithm yields several connected regions. Some of the blocks are discarded since they are at corners of the scene. Hence, there were only two blocks of region. To extract the contour features, boundary points are determined by tracing neighboring pixels of the connected region. Next, the perimeter and area are calculated to obtain the *CR*. Furthermore, we also used chain coding to get the angles and the curve segment length for the curvature of a certain connected block. Finally, the bending energy is calculated using (5). Fig. 1(c) and (d) show the detected clutter and target based on these two features. The circularity ratio is found to be 0.14 and 1 for clutter and target, respectively. Whereas the bending energy is 0.23 and 1.75 for clutter and target, respectively. The classifier then successfully detects the non-linear clutter and the point target in the radar scene.

## V. CONCLUSION

A novel technique based on multiple contour features is presented in this paper. The simulation result suggests that the proposed method is effective in the presence of elliptical clutter. The method is also applicable when the target merges with some of the pixels of this type of inhomogeneous complex shaped clutter.